\documentclass{article}
\input epsf.sty
\input colordvi.sty
\usepackage[dvips]{graphicx}
\newcommand{\bright}{\begin{flushright}}
\newcommand{\eright}{\end{flushright}}
\newcommand{\bminip}{\begin{minipage}}
\newcommand{\eminip}{\end{minipage}}
\newcommand{\bcent}{\begin{center}}
\newcommand{\ecent}{\end{center}}
\newcommand{\beq}{\begin{equation}}
\newcommand{\eeq}{\end{equation}}
\newcommand{\beqa}{\begin{eqnarray}}
\newcommand{\eeqa}{\end{eqnarray}}
\newcommand{\barr}{\begin{array}}
\newcommand{\earr}{\end{array}}

\newcommand{\reflef}{(\ref}

\newcommand{\Lmd}{\Lambda}

\newcommand{\lsim}{\mbox{\raisebox{-.3em}{$\;\stackrel{<}{\sim}\;$}}}

\begin{document}
\baselineskip=0.6cm

\bcent
{\LARGE\bf Revised fits to $\Delta\alpha/\alpha$ in consistency with the accelerating universe}\\[.8em]
{\large Yasunori Fujii}\\
Advanced Research Institute for Science and Engineering, \\[.0em]
Waseda University, 169-8555 Tokyo, Japan\\[.0em]
\ecent
\mbox{}\\[-1.6em]

\bminip{14.5cm}
{\large\bf Abstract}\\[-.7em]

\hspace*{1em}An attempt is made for a new type of analysis of the time-variability of the fine-structure constant trying to fit the most recent result from the laboratory measurements, the Oklo constraint and the data from the QSO absorption lines all in consistency with the accelerating universe.
\eminip
\mbox{}\\[1.0em]

We have developed a theoretical model of the accelerating universe based on the scalar-tensor theory \cite{cup}-\cite{YFNaha}.  We found that the simplest version of the scalar-tensor theory with an assumed cosmological constant $\Lmd$ included must be modified in order for the scalar field, serving as the dark energy,  to provide with its energy density which behaves as a plateau acting like a temporary constant to cause a mini-inflation, or an acceleration of the universe.  For this purpose we introduced another scalar field $\chi$ providing a temporary potential which traps the original scalar field $\sigma$, thus implementing the scenario of a decaying cosmological constant; $\Lmd \sim t^{-2}$ in the Einstein conformal frame, realized numerically by $10^{-120}\sim (10^{60})^{-2}$ in the reduced Planckian unit system, with $t_0\sim 10^{60}$ for the present age of the universe expressed in units of the Planck time $\sim 10^{-43}{\rm s}$.  This leaves the ``fine-tuning problem" solved and the ``coincidence problem" eased, at least \cite{cup,PTPInv,YFNaha}.

During this trapping process, we naturally expect an oscillatory behavior $\sigma (t)$ as a function of the cosmic time $t$.  We also showed that this might be observed as a time-dependent fine-structure constant.  In \cite{plb}, we tried to fit the most recent results obtained from QSO absorption spectra \cite{lev1}-\cite{porsev}, the constraint from the Oklo phenomenon \cite{oklo}-\cite{Gould} and the laboratory measurements of atomic clocks of Yb$^+$ and Hg$^+$ \cite{peik} in terms of the theoretical curves chosen to fit the observed cosmological acceleration.

We obtained a class of fits which we considered to be a reasonable success of the theoretical model \cite{plb}.  We point out that the achieved strong constraint is due to the unique analysis of Single Ion Differential $\alpha$ Measurement (SIDAM) featuring  the measurement at individual redshift \cite{lev1}-\cite{porsev}, which is better suited for comparison with the theoretical assumption of the oscillatory behavior, than other approaches  making extensive use of the averaging processes \cite{murph}-\cite{chand}, for which   the  phenomenological analyses tended to produce less constrained results \cite{plb2}.


More important to be noticed currently, however, we have come to know a latest analysis of the atomic clocks of Al$^+$ and Hg$^+$  giving the result \cite{nist}; $(\dot{\alpha}/\alpha)_0 =(-1.6 \pm 2.3)\times 10^{-17}{\rm y}^{-1}$, which is about an order of magnitude smaller than the previous result \cite{peik}, and is translated into
\beq
y_0'=\left( \frac{d}{dz}\frac{\Delta\alpha}{\alpha} \right)_0 =-H_0^{-1}\left( \frac{\dot{\alpha}}{\alpha} \right)_0=(2.1\pm 3.1)\times 10^{-7},
\label{horz_1}
\eeq
where $z$ is redshift and the subscript 0 implies the present time.

By extrapolating  $\dot{\alpha}/\alpha \sim 10^{-17}{\rm y}^{-1}$ assumed to be time-independent, at this moment, back to the epochs $\sim 10^{10}{\rm y}$ ago we find $\Delta\alpha/\alpha$ as small as $\sim 10^{-7}$.  Theoretically, however, today's small value by no means implies uniquely the same at earlier epochs \cite{plb,plb2}.  In the analysis of \cite{plb}, on the other hand,  rather large values of the QSO results \cite{lev1}-\cite{porsev} can be fitted only by $y'_0$ so large compared with \reflef{horz_1}).   It seems as if we have come to the extent that we may no longer be allowed to identify both of today and the Oklo time with the same type of zeros of $\Delta\alpha/\alpha$.  In spite of certain uncertainties in \cite{lev1}-\cite{porsev}, as well as other results \cite{murph,chand}, together with still the preliminary aspect admitted in \cite{nist},  it seems appropriate and urgent to show if an alternative analysis can be applied basically to the same pursuit in terms of the oscillating $\alpha$ and SIDAM, as in \cite{plb}.

\mbox{}\\[-8.6em]
\bminip{19cm}
\hspace*{2.0em}
\includegraphics[width=.65\textwidth]{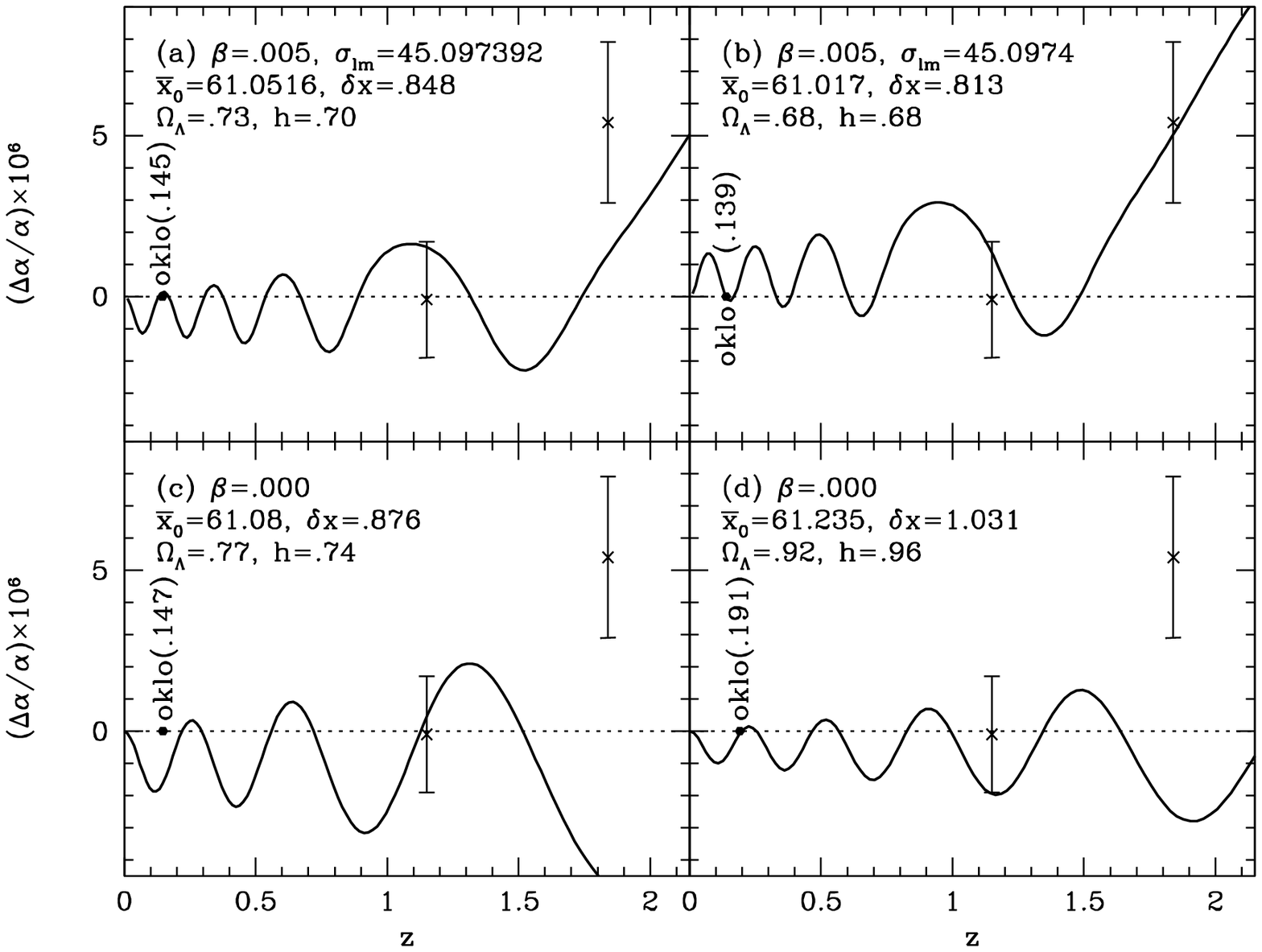}
\eminip
\mbox{}\\[-2.5em]
\hspace*{3em}
\begin{figure}[h]
\caption{Typical plots of the theoretical curve for $(\Delta\alpha/\alpha)\times 10^6$ as a function of redshift $z$.  The Oklo phenomenon having occurred $\approx 1.95\times 10^9{\rm y}$ ago corresponds to $z\sim 0.15$, indicated in the parenthesis of each plot, depending on how the scale factor varies with the cosmic time.  The two QSO data \cite{lev1}-\cite{porsev} are shown; $-0.12\pm1.79$ and $5.66\pm2.67$ for $z=1.15$ and $1.84$, respectively.  At the present time $z=0$ the portion of the ``flat" top, or the bottom, is hardly discernible, though it may extend over the time-span as long as $\sim 10^6{\rm y}$, as will be shown toward the end of this article.  We commonly choose the initial values at $t_1=10^{10}$ in the reduced Planckian unit system, as in Fig. 5.8 of \cite{cup}; $\sigma_1=6.77341501, \sigma'_1=0, \chi_1=0.21, \chi'_1=0$, where the prime is for the derivative with respect to $\tau=\ln t$.  The lower plots (c) and (d) with $\beta=0$ are for the potential in the original theoretical model, while the upper plots (a) and (b) are  for the Sine-Gordon part of the potential corrected by \reflef{horz_3}) with $\beta=0.005, \Gamma =1.2\times 10^{-4}$ also with $\sigma_{\rm lm}$ as indicated in the diagrams.  We applied the technique of shifting the evolution variable $x$ as explained in the text on $\bar{x}$ and $\delta x$.  In (c) we cannot make the second zero of oscillation to be sufficiently close to the Oklo time as far as $\Omega_\Lmd$ and $h$ remain close to the observed values $\Omega_\Lmd =0.72\pm 0.06$ and $h=0.73\pm 0.03$ \cite{accu,wmap}, while we manage it in (d) at the cost of unacceptably larger values for them, for the reasons we explain in the text.  In the plots (a) and (b), in contrast, with the potential  modified phenomenologically we achieve satisfactory agreements with the observational results.
}
\end{figure}

From this point of view, we now assume that we at the present epoch happen to be at the top or the bottom of the oscillation, with the Oklo time simply sharing  nearly the same height as of today, as illustrated in the diagrams (a) and (b) in Fig. 1.  Choosing $y'_0=0$ might be accepted at least as a first approximation  to deal with the laboratory measurements as small as $\sim 10^{-17}{\rm y}^{-1}$ \cite{nist}.  This approach, however, requires the separation $\Delta t_{\rm oklo} \approx 1.95\times 10^9{\rm y}$ to the Oklo phenomenon to be close to one ``period" of oscillation.  Conversely, the period must be nearly as small as $\Delta t_{\rm oklo}$.  Comparing this with the situation in \cite{plb} in which the separation between two zeros is not related directly to the period, or the half-period, of the oscillation by adjusting the value of $y'_0$, we find ourselves obviously more constrained.  This poses in fact a non-trivial task as will be also illustrated in (c) and (d) as long as we stay in the conventional way of understanding the accelerating universe.

\mbox{}\\[6.0em]
\bminip{19cm}
\mbox{}\\[-5.5em]
\hspace*{8.5em}
\includegraphics[width=.55\textwidth]{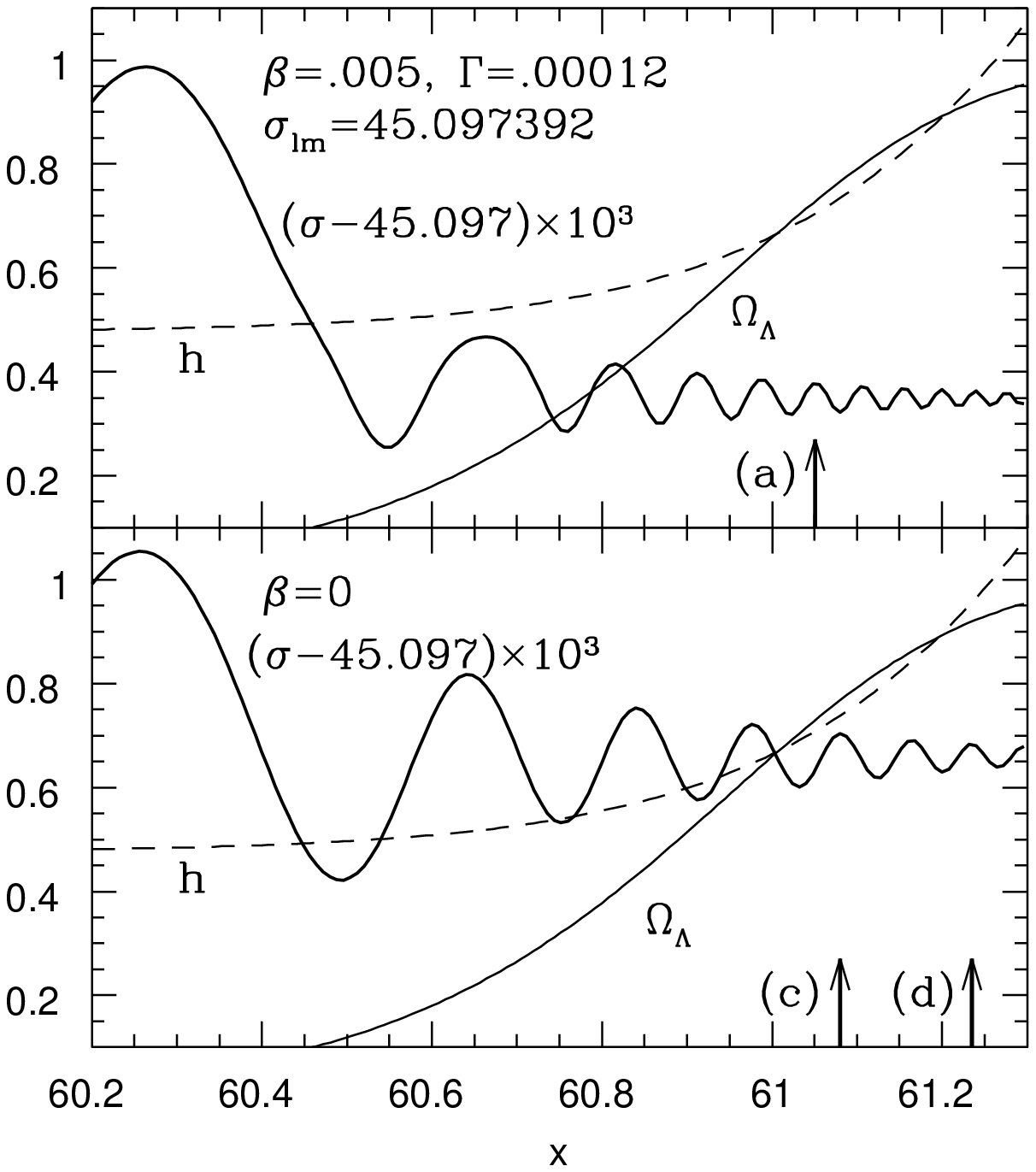}
\eminip
\mbox{}\\[-2.5em]
\begin{figure}[h]
\caption{
Computed values of $\sigma, \Omega_\Lmd$ and $h$ are plotted against the evolution coordinate $x=\log_{10}t$ for the modified and the unmodified potentials in the upper and the lower panels, respectively.  The same initial values as in Fig. 1 were used in obtaining the cosmological solutions.   We may re-interpret $x$ also as the shifted coordinate $\bar{x}=x+\delta x$.  Today's age of the universe $t_0=1.37\times 10^{10}{\rm y}$ corresponds to $x_0=60.204$.  The arrows marked with the names of plots in Fig. 1 indicate the values $\bar{x}_0$ which corresponds to today.  $\Omega_\Lmd$ is simply computed in the Einstein conformal frame at $\bar{x}_0$ for $x$, while $h$ is today's value of $H$ in units of $H_{100}\equiv 100$ km/sec/Mpc $=0.875\times 10^{-60}$ in the reduced Planckian units, and is estimated by $h=b'(\bar{x}_0)t_0^{-1}/H_{100}$, where $b'=(d/d\tau)(\ln a)$.   The values of parameters in \reflef{horz_3}) are also indicated in the upper panel.
}
\end{figure}

Let us illustrate how the scalar field $\sigma$ develops with time around the present epoch toward the end of mini-inflation.  We will also argue how we may re-use the existing solutions of the cosmological equation which is highly nonlinear.  In the lower panel of Fig. 2, we plot $\sigma$ as a function of the evolution coordinate $x=\log_{10}t$ with the initial value $\sigma_1 =6.77341501$, somewhat different from that used in Fig. 5.8 of \cite{cup}, but with others remaining the same, imposed at $x_1=10$ in the reduced Planckian units, so that we find $\Omega_\Lmd$ and $h$ somewhat smaller than the accepted values $0.72\pm 0.06$ and $0.73\pm 0.03$ \cite{accu,wmap}, respectively, at $x_0=60.204$ corresponding to today $t_0\approx 1.37\times 10^{10}{\rm y}$.  We may interpret $x$ in Fig. 2 not only to be a real evolution coordinate but also as being``shifted" by $\bar{x}=x+\delta x$.

In the lower panel of Fig. 2 we pick up a peak at  $x=61.08$, for example, chosen for $\bar{x}_0$ corresponding to today, as indicated by the arrow marked with  (c).  We then have $\delta x=61.08-60.204=0.876$, also allowing us to assign the reasonable set of values $\Omega_\Lmd\sim 0.77$ and $h\sim 0.74$.  The initial time $t_0-t_1$ ago with $t_1=10^{10}$ when the initial conditions were given originally can be re-interpreted as the  occurrence at $\bar{t}_0-t_1 =(t_0-t_1) +\delta t$ ago, where $\delta t/t\approx \delta x\ln 10$.  This is what we mean by re-using the existing solutions.  Shifting the initial time may be viewed as another way of revising initial conditions.

The above choice of $\bar{x}_0$ has been made in such a way that it corresponds precisely to one of the tops (or the bottoms) of the curve of $\sigma$, hence to the same of $\Delta\alpha/\alpha$ computed according to
\beq
\frac{\Delta\alpha}{\alpha}={\cal Z}\frac{\alpha}{\pi}\zeta\Delta\sigma \approx 1.84\times 10^{-2}\Delta\sigma,
\label{horz_2}
\eeq
which is 4 times larger than in (1) of \cite{plb} due to our using more realistic half-spin particles rather than spinless particles in the toy model of \cite{cup,plb}, included in the photon self-energy part, but using the same regularization method in terms of the continuous spacetime dimension.

From this ``top," chosen to be a zero of $\Delta\alpha$ for today, we draw a horizontal line to the right now in (c) of Fig. 1.  The near crossing with the curve of $\Delta\alpha/\alpha$ is supposed to coincide with the Oklo.  Obviously we fail because the period of oscillation is too long.  We may improve the situation by exploiting the observation in Fig. 2 showing that the $\sigma$ oscillation ``dwindles" with time in the amplitude as well as in the period.  Choose another peak at $\bar{x}_0=61.235$, for example, further to the right in the lower panel of Fig. 2, as shown by the arrow marked with (d).  Repeating the same procedure as above, we now have a shorter period to cross the Oklo point as in (d) of Fig. 1, but with unacceptably larger values of $\Omega_\Lmd$ and $h$ as a trade-off.  This type of failure appears to be rooted more generally, because from a matured ``hesitation" behavior the less matter component and the faster growing scale factor should follow, as we can easily recognize in Fig. 5.8 of \cite{cup}.

This discussion also suggests a possible way out, faster dwindling of the oscillation as illustrated in the upper panel of Fig. 2, perhaps due to the potential  ``sharpened" toward its bottom.  The desired potential, partially of the Sine-Gordon type in the original model, is modified according to the assumed  phenomenological parametrization defined by
\beq
\sin (\kappa\sigma) \rightarrow \sin (\kappa\sigma) -\beta\exp \left[ -\left( \frac{\sigma -\sigma_{\rm lm}}{\Gamma} \right)^2 \right],
\label{horz_3}
\eeq

The parameter $\Gamma$ in the Gaussian distribution has been chosen to be of the same order of magnitude as the range $\lsim 10^{-4}$ across which $\sigma$ changes around the present time, as we find in the lower panel of Fig. 2.  The limiting value $\sigma_{\rm lm}$ may be defined in principle by $\kappa\sigma_{\rm lm}=(3\pi/2)$ (mod $2\pi$), but appears slightly different in practice.  We consider its local value to be another adjustable parameter.   The overall factor $\beta$ has been chosen to be $0.005$, indicating that the required change of depth of the potential is only about 0.5\%.  With a choice of $\Gamma=1.2\times 10^{-4},\ \sigma_{\rm lm}=45.097392$, we pick up the peak at $\bar{x}_0=61.0516$, as indicated by the arrow in the upper panel of Fig. 2, thus yielding the plot (a) of Fig. 1.  Note that $\Omega_\Lmd$ and $h$ turn out to fall into the reasonable range caused by a relatively minor difference between the behaviors of $\sigma$ in the upper and lower panels of Fig. 2.  We point out that the fit results in no significant difference in the cosmological behavior of the solution, including the equation of state for the dark energy remaining close to $-1$, basically the same as in Fig. 6 of \cite{plb}.

A similar fit is also shown in the plot (b) of the same Fig. 1.  This, though with a somewhat smaller $h$, together with (a) represents our goal in the present article illustrating how successful we can be in reconciling the small laboratory value with the large QSO results.  It might be worth pointing  out that it would have been rather easier to find solutions with smaller oscillations, like in the plot (f) of Fig. 1 of \cite{plb}, for example, if the QSO measurements were to yield smaller values.  At the same time, however, we add that $\Delta\alpha/\alpha$ could be considerably larger at earlier epochs, like those of CMB or the primordial nucleosynthesis, as inferred from Fig. 5.10 of \cite{cup} or Figs. 4 and 5 of \cite{plb}.

We are still not sure if the potential modified above only slightly through the phenomenological parameters can be derived by revising some of the parameters and initial values of the original theory, leaving wider searches for solutions as future tasks.  We do not know either why we have $y'_0=0$ theoretically at the present epoch. In spite of these questions, we still believe the present attempt to be important from wider views.    It is also remarkable, on the other hand, to find that the measurements at the present epoch have reached the level to affect the assumed earlier temporal behavior of the fine-structure constant in a non-trivial manner.

After all of these discussions comes an obvious message.  For further development of the issue, it is desperately necessary to start new analyses at least in some of the epochs in which no observation or experiment has been available yet with sufficient accuracy.  See \cite{yfai} for an example of insufficient rigor of the argument on the constraint expected from the decay $^{187}{\rm Re}\rightarrow ^{187}\hspace{-.3em}{\rm Os}$. 
\\[1.0em]

We add a few related comments.  We first notice that the correction term \reflef{horz_3}) applies only to a specific cycle of the potential.  We may find the similar term which applies to any other cycles for a given phase.  Let us start with the argument on $\sigma_{\rm lm}$ mentioned  after \reflef{horz_3}), writing
\beq
\kappa\sigma_{\rm lm}=2\pi \frac{3}{4}(1+\delta)\quad (\mbox{mod}\ 2\pi).
\label{horz_4}
\eeq
For the value $\sigma_{\rm lm}$ indicated in the plot (a) of Fig. 1 with $\kappa=10$, we find $\delta =0.032972$.  We then write
\beq
\kappa\sigma =\kappa (\sigma -\sigma_{\rm lm})+\kappa\sigma_{\rm lm} \equiv v+\frac{3\pi}{2}(1+\delta),
\label{horz_5}
\eeq
where $|v|\ll 1$ can be a good approximation.  We thus derive
\beq
-\left( \frac{\sigma -\sigma_{\rm lm}}{\Gamma} \right)^2\approx -2\frac{1+\sin (\kappa\sigma -(3\pi\delta/2))}{(\kappa\Gamma)^2},
\label{horz_6}
\eeq
which can be substituted into \reflef{horz_3}) for an expression to be applied to any portion of the Sine-Gordon potential, as far as we can use a common value of $\delta$.  Unfortunately, however, examining the earlier mini-inflation to have occurred around $x=26$ in the solution used in (a) of Fig. 1, for example, we find $\delta =-0.0902$, quite different from the above value determined around today.  We suspect that $\sigma$ may behave differently depending on how it was last trapped by the potential well.    In the solution under the current study we have only two occurrences of mini-inflation.  In all the other cycles, $\sigma$ flies so high above the potential that no details near the minima affect the overall behavior of $\sigma$.

In Fig. 1 the Oklo constraints were indicated simply by small filled circles, but the real error bars  are much smaller, nearly invisible.  More exact numerical estimates on the fits (a) and (b) in Fig. 1 result in $0.12\pm 0.01$ and $0.006\pm 0.071$, respectively, for $(\Delta\alpha/\alpha) \times 10^6$ during the time interval of $\sim 0.1\times 10^9$ years as an interpretation of $\Delta t_{\rm oklo}=(1.95\pm 0.05) \times 10^9{\rm y}$ \cite{neudet}.  We refrain from further improving the accuracy of the calculation because the present approach does appear to allow its implementation, also in view of the current theoretical limitation of the phenomenological conclusion from the Oklo phenomenon, as will be sketched below.

We have the upper bound $|\Delta\alpha/\alpha| \lsim 10^{-8}$ by the conventional Coulomb-only estimate \cite{DD,yfoklo}, by ignoring the effect of the strong interaction.  According to the argument exploiting the near cancellation between the two contributions, due to the Coulomb and the strong interactions as elaborated in Appendix A of \cite{ptb},\footnote{The second term on the far right-hand side of (A.3) and on the right-hand side of the second equation of (A.4) might be better multiplied by $n$ for possible non-perturbative dependence on $\alpha_s$, replacing $\xi$ in (A.9)-(A.11) effectively by $n\xi$. } however, the process might be interpreted to produce the result which lies, roughly speaking, in the range an order of magnitude larger or smaller, representing the size $\sim 0.1-0.001$ in the scale of Fig. 1, hence  leaving the present analysis nearly intact.  Model dependencies of this type, including asymptotic freedom in QCD hampering the calculation applied to the small energy-scale for atomic transitions, are unavoidable before we find more reliable way to relate the two contributions to each other.

It seems useful to look at a magnified view of the top (or the bottom) portion of $\Delta\alpha/\alpha$ near the present time.   An example is shown in Fig. 3 for the plot (a) of Fig. 1.  The curve is nearly a symmetric parabola assumed centered at $x_0=60.204$ for computational simplicity.  At $x-x_0 =\pm 1.27\times 10^{-4}$, we reach the gradient $dy/dx=\pm 7.29\times 10^{-7}$, which translates into $dy/dz =\pm 3.1 \times 10^{-7}$ corresponding to both ends of \reflef{horz_1}).  The above interval $ x-x_0$ translates also into the time interval $ t-t_0 \approx t_0 ( x-x_0)\ln 10 =\pm 4.0\times 10^6{\rm y}$.  This is too short to be seen in Fig. 1, yet too long to wait until we can detect any change, if any, of the gradient even with the accuracy of \cite{nist}.\\[-4.0em]

\mbox{}\\[-2.0em]
\bminip{15cm}
\mbox{}\\[-10.5em]
\hspace*{8.5em}
\includegraphics[width=.65\textwidth]{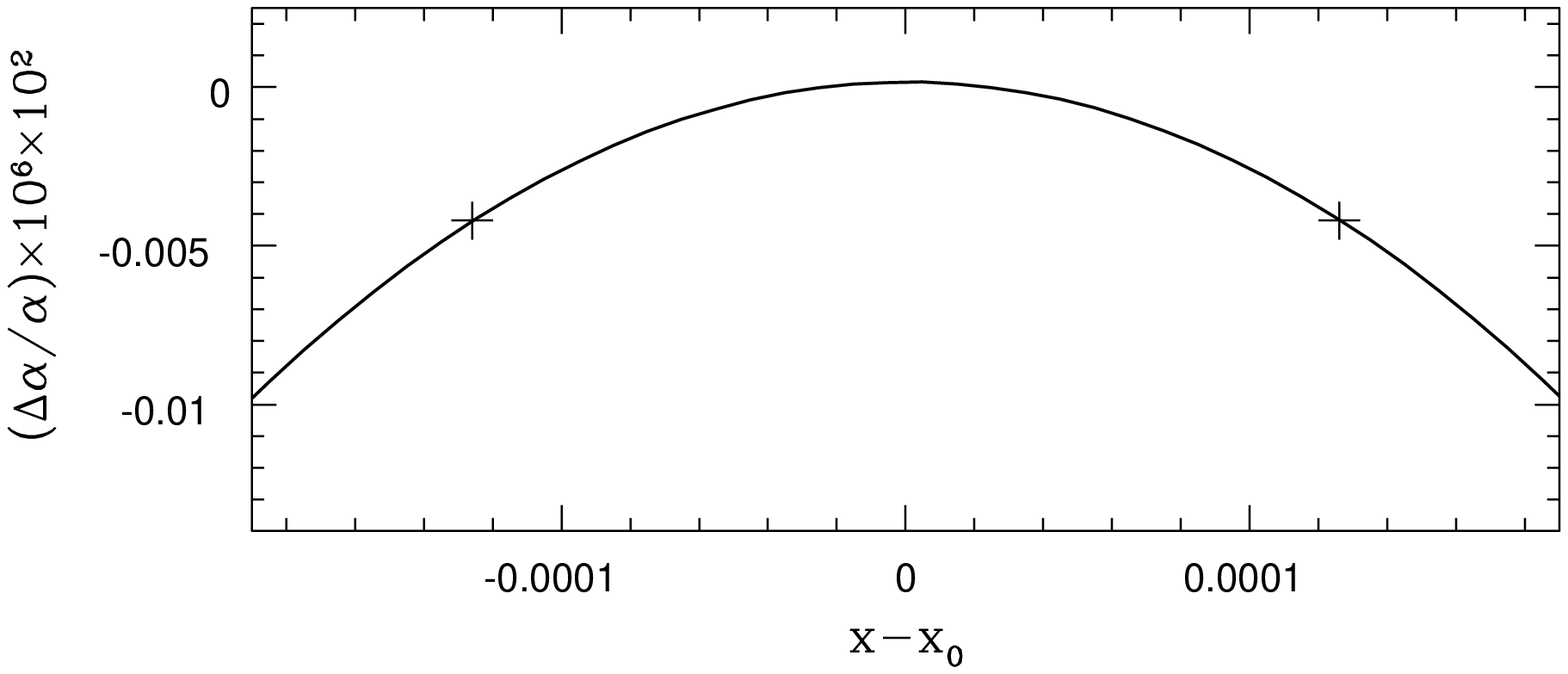}
\eminip
\mbox{}\\[-2.9em]
\begin{figure}[h]
\caption{Magnified view of the top portion of $\Delta\alpha/\alpha=y$ in the vicinity of today $x_0=60.204$ for the plot (a) of Fig. 1.  The limiting values of the gradient $\pm 3.1\times 10^{-7}$ in \reflef{horz_1}) correspond to $dy/dx =\pm 7.29\times 10^{-7}$ at $x-x_0 =\pm 0.000127$, or $t-t_0 =\pm 4\times 10^6{\rm y}$, as indicated by the crosses.
}
\end{figure}

\noindent
{\Large\bf Acknowledgements}

I would like to thank Hidetoshi Katori and Paolo Molaro for their valuable discussions.



\begin{thebibliography}{9}


\bibitem{cup}Y. Fujii and K. Maeda, {\it The scalar-tensor theory of gravitation}, Cambridge University Press, 2003.
\bibitem{yfbls}Y. Fujii, in {\it Gravity, Astrophysics and Strings at the Black Sea}, St. Kliment Ohridski University Press, 2003.
\bibitem{PTPInv}Y. Fujii, Prog. Theor. Phys. {\bf 118} (2007), 983.
\bibitem{YFNaha}Y. Fujii, Proc. Workshop on Cold Antimatter Plasmas and Application to Fundamental Physics, Feb. 20-22, 2008, Naha, Japan [gr-qc/0803.3103].
\bibitem{plb}Y. Fujii, Phys. Lett. {\bf B660} (2008), 87.
\bibitem{lev1}S. Levshakov, P. Molaro, S. Lopez, S. D'Odorico, M. Centuri\'{o}n, P. Bonifacio, I. Agafonova, and D. Reimers, A\&A, {\bf 466} (2007), 1077.
\bibitem{lev2}S. Levshakov, I. Agafonova, P. Molaro, and D. Reimers, submitted to A\&A, 2007.
\bibitem{lev3}P. Molaro, D. Reimers, I. Agafonova, and S. Levshakov, to appear in Proc. {\it Atomic clocks and fundamental constants}, ACFC 2007, Bad Honnef, June 2007.
\bibitem{porsev}S.G. Porsev, K.V. Koshelef, I.I. Tupitsyn, M.G. Kozlev, D. Reimers and S.A. Levshakov, Phys, Rev. {\bf A76} (2007), 052507.
\bibitem{oklo}A.I. Shlyakhter, Nature {\bf 264} (1976), 340; ATOMKI Report A/1 (1983), unpublished [physics/0307023]. 
\bibitem{DD}T. Damour and F. Dyson, Nucl. Phys. {\bf B480} (1996), 37. 
\bibitem{yfoklo}Y. Fujii, A. Iwamoto, T. Fukahori, T. Ohnuki, M. Nakagawa, H. Hidaka, Y. Oura, and P. M\"{o}ller, Nucl. Phys. {\bf B573} (2000), 377.
\bibitem{ptb}Y. Fujii, in Astrophysics, clocks and fundamental constants, Lecture Notes in Physics, {\bf 648}, eds. S.G. Karschenboim and E. Peik, Springer, 2004.
\bibitem{Gould}Yu.V. Petrov, A.I. Nazarov, M.S. Onegin, V.Yu. Petrov and E.G. Sakhnovsky, Phys. Rev. C {\bf 74} (2006), 064610.  C.R. Gould, E.I. Sharapov and S.K. Lamoreaux, Phys. Rev. C {\bf 74} (2006), 024607.
\bibitem{peik}E. Peik, B. Lipphardt, H. Schantz, Chr. Tamm, S. Weyers, and R. Wynands, Proc. 11th Marcel Grossmann Meeting, 2006 [physics/0611088].
\bibitem{murph}M.T. Murphy, V.V. Flambaum, J.K. Webb, J.X. Prochaska, and A.M. Wolfe, in Astrophysics, clocks and fundamental constants, Lecture Notes in Physics, {\bf 648}, eds. S.G. Karschenboim and E. Peik, Springer, 2004.  M.T. Murphy,  J.K. Webb and V.V. Flambaum, Phys. Rev. Lett. {\bf 99} (2007), 239001; MNRAS {\bf 384} (2008), 1053.
\bibitem{chand}H. Chand, R. Srianand, P. Petitjean, and B. Aracil, A\&A, {\bf 417} (2004), 853; R. Srianand, H. Chand, P. Petitjean, and B. Aracil, Phys. Rev. Lett. {\bf 92} (2004), 121302.
\bibitem{plb2}Y. Fujii and S. Mizuno, Int. J. Mod. Phys. {\bf D14} (2005), 677.  Y. Fujii, Phys. Lett. {\bf B616} (2005), 141.
\bibitem{nist}T. Rosenband, D.B. Hume, P.O. Schmidt, C.W. Chou, A. Brusch, L. Lorini, W.H. Oskay, R.E. Drullinger, T.M. Fortier, J.E. Stalnaker, S.A. Diddams, S.A. Swann, N.R. Newbury, W.M. Itano, D.J. Wineland, and J.C. Bergquist, Science {\bf 319} (2008), 1808.
\bibitem{accu}A.G. Riess et al., Astron. J. {\bf 116} (1998), 1009. S. Perlmutter et al., Nature {\bf 391} (1998), 51; Astrophys. J. {\bf 517} (1999), 565.
\bibitem{wmap}D.N. Spergel et al., Astrophys. J. Suppl. Ser. {\bf 170} (2007), 377.
\bibitem{yfai}Y. Fujii and A. Iwamoto, Phys. Rev. Lett. {\bf 91} (2003), 261101; Mod. Phys. Lett. {\bf A20} (2005), 2417.
\bibitem{neudet}R. Neudet, {\it Oklo: des r\'{e}acteurs nucl\'{e}aires fossiles, Collection CEA} (Eyrolles. Paris, 1991). H. Hidaka, Radioisotopes {\bf 46} (1997), 96.


\end{thebibliography}
\end{document}